\def\aj{{AJ}}

\def\apj{{ApJ}}

\def\mnras{{MNRAS}}

\documentclass[cup5b]{caps}

\usepackage{graphicx}

\begin{document}

\pagenumbering{arabic}

\author[]{Peter Erwin$^{1}$, Alister W. Graham$^{2}$, Nicola Caon$^{1}$ \\
(1) Instituto de Astrof\'{\i}sica de Canarias, La Laguna, Spain \\
(2) Department of Astronomy, University of Florida, Gainesville, FL, 
USA}

\chapter{The Correlation Between Supermassive Black Hole Mass and the
Structure of Ellipticals and Bulges}

\begin{abstract}

We demonstrate a strong correlation between supermassive black hole
(SMBH) mass and the global structure of ellipticals and bulges: more
centrally concentrated bulges and ellipticals (higher S\'ersic index
$n$) host higher-mass black holes.  This correlation is as good as
that previously found between SMBH mass and central velocity
dispersion, with comparable scatter.  In addition, by carefully
modeling the bulges of disk galaxies so that bars, inner disks, and
the like do not accidentally contribute to bulge light, we find that
the correlation between SMBH mass and bulge or elliptical luminosity
is similarly close.

\end{abstract}

\section{Introduction}

Observations now show that supermassive black holes (SMBHs; $M_{\rm
BH} \sim 10^{6}$-Ð$10^{9} M_{\odot}$) are probably present at the
centers of most, if not all, elliptical galaxies and bulges
(collectively, ``bulges'').  More recent studies found a strong
correlation between SMBH mass and the central stellar velocity
dispersion $\sigma_{0}$ of bulges (Ferrarese \& Merritt 2000; Gebhardt
et al.\ 2000), much stronger than the previous correlation found
between bulge luminosity and SMBH mass (Magorrian et al.\ 1998).

Graham, Trujillo, \& Caon (2001a) have shown that the central
concentration of bulge light $C_{re}$, measured within the half-light
radius $r_{e}$, positively correlates with $\sigma_{0}$ of the bulge. 
In addition, Graham (2002b) found that the S\'ersic (1968) index $n$,
determined from $r^{1/n}$ fits to bulge light profiles, correlated
extremely well with $\sigma_{0}$.  Taken together, this suggests that
there may be a correlation between SMBH mass and the global
\textit{structure} of bulges.  Here, we demonstrate that such a
correlation does indeed exist: more concentrated bulges (higher Sersic
index $n$) have more massive SMBHs.  (A preliminary version of this
correlation, using $C_{re}$, was presented in Graham et al.\ 2001b). 
This correlation is as strong as that between SMBH mass and stellar
velocity dispersion, and has comparable scatter.

By taking care in isolating and accurately modeling the bulges of disk
galaxies, it is possible to make more accurate estimates of bulge
luminosity.  This lets us re-evaluate the relation between bulge
luminosity and SMBH mass and avoid scatter introduced by, e.g.,
assigning a fixed fraction of a disk galaxy's light to the bulge based
solely on its Hubble type.  The result is that the correlation between
$R$-band luminosity and SMBH mass proves to be almost as strong as the
$\sigma_{0}$-Ð$M_{\rm BH}$ and $n$-Ð$M_{\rm BH}$ correlations,
with similar scatter.

\section{Data and Analysis}

\begin{figure}

  \begin{center}
  \includegraphics[scale=0.65]{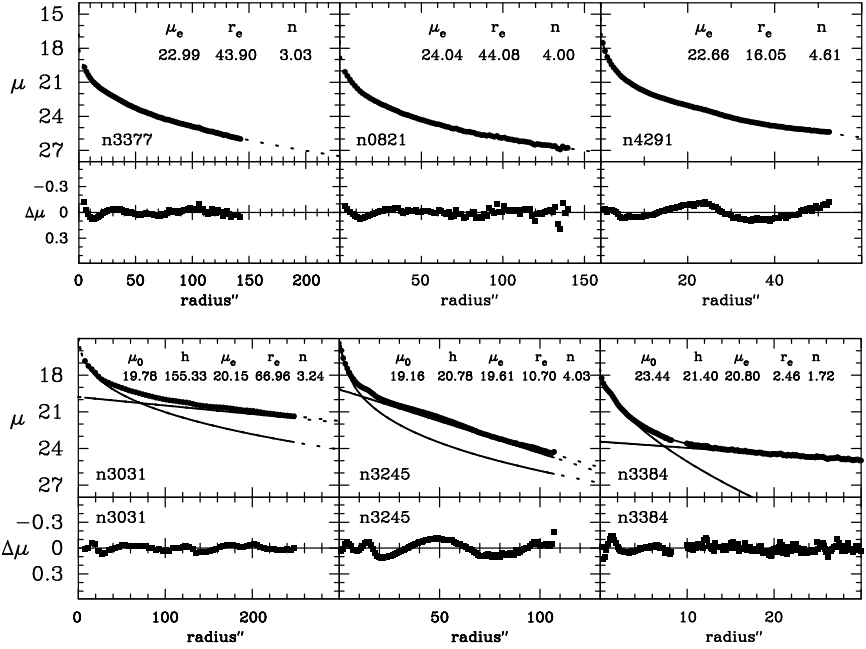}
  \caption{Sample fits to galaxy profiles.  Elliptical galaxies (top)
  are fit by a S\'ersic $r^{1/n}$ model; disk-galaxy profiles (bottom)
  are fit with a S\'ersic + exponential model.  All fits incorporate
  seeing convolution.  Most profiles are from ellipse fits; the
  NGC~3384 profile is a major-axis cut, almost perpendicular to the
  bar.} \label{fig:erwin-fits}
  \end{center}
\end{figure}

We searched the public archives for $R$-band galaxy images which were
large enough to guarantee good sky subtraction (galaxy well within the
field of view) and which had no central saturation.  Most of the
images came from the Isaac Newton Group and \textit{Hubble Space
Telescope} (\textit{HST}) archives; we used some $I$-band (F814W)
images from \textit{HST} if there were no $R$-band (F702W) images
available.  We found useful images for 21 of the 30 galaxies with
``reliable'' SMBH measurements (Table~1 in Merrit \& Ferrarese 2001). 
The galaxy isophotes of the reduced, sky-subtracted images were fit
with ellipses.  We then fit the resulting major-axis surface
brightness profiles with seeing-convolved S\'ersic $r^{1/n}$ models
for the ellipticals and a combined, seeing-convolved exponential disk
+ S\'ersic bulge model for the disk galaxies
(Figure~\ref{fig:erwin-fits}).  In several cases, fitting the global
profile of disk galaxies --- particularly when strong bars are present
--- can badly mismeasure the bulge; special care was needed in those
cases (see Bulge-Disk Decomposition, below).  We excluded the inner
$\sim 100$ pc from the fits, to avoid contamination of the profiles by
power-law cores, stellar or active nuclei, and nuclear disks.

Black hole masses and central velocity dispersions were taken from the
compilation of Merritt \& Ferrarese (2001), with updated black hole
masses from Gebhardt et al.\ (2002).  We used aperture photometry from
the literature to photometrically calibrate $R$-band profiles for 13
galaxies.  Bulge absolute magnitudes for these galaxies were then
calculated using the best-fit S\'ersic model and surface-brightness
fluctuations distances from Tonry et al.\ (2001) or kinematic
distances from LEDA (with $H_{0} = 75$).  The central concentration
$C_{re}$ was calculated from the S\'ersic index $n$ (Graham et al.\
2001a); preliminary results using $C_{re}$ were presented in Graham et
al.\ 2001b.

\subsection{Bulge-Disk Decomposition; Distinguishing Ellipticals
from S0's}

\begin{figure}
  \centering
\includegraphics[scale=0.65]{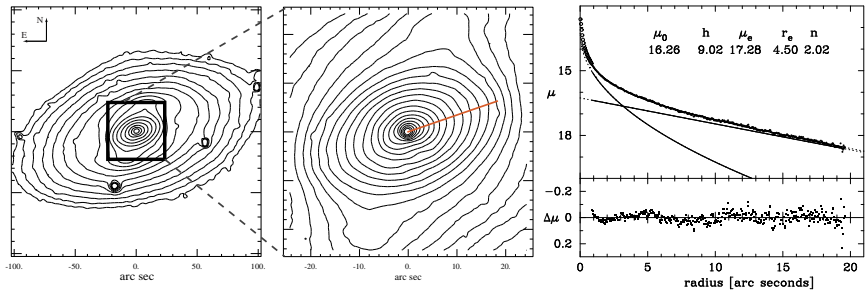}
  \caption{An example of careful bulge-disk decomposition: the
  isophotes inside the bar of NGC~2787 appear at first glance to be a
  bulge, but $\sim 2/3$ of the light is actually from an exponential
  inner disk, distinct from the outer disk (Erwin et al.\ 2003). 
  Here, $R$-band isophotes show the inner-disk + bulge in context,
  along with an \textit{HST} profile from a major-axis cut (diagonal red line
  in middle panel) and our S\'ersic + exponential fit (right panel).}
  \label{fig:erwin-n2787}
\end{figure}

Critical to determining both the structure and luminosity of bulges in
disk galaxies is the proper separation of the bulge from other galaxy
components: not just from the main disk, but also from bars, lenses,
and inner disks.  We made a careful analysis of each disk galaxy,
identifying cases where global bulge-disk decomposition would not
work.  In some galaxies, the existence of a lens or inner disk allowed
us to make exponential + bulge decompositions without using the outer
disk (e.g., NGC~2787 in Figure~\ref{fig:erwin-n2787}; see Erwin et
al.\ 2003 for a detailed discussion of this galaxy, including how a
global bulge-disk decomposition mismeasures the bulge).  For the Milky
Way, we used the perpendicular near-IR profile of Kent, Dame, \& Fazio
1991 as the bulge profile.

Equally important is identifying disk galaxies which have been
misclassified as ellipticals, since otherwise a disk + bulge structure
will be erroneously described with a single (S\'ersic) model.  Using
information from the literature, including the kinematic study by Rix,
Carollo, \& Freeman (1999), along with morphological and photometric
information, we found that NGC~2778 and NGC~4564 are probably S0
galaxies, despite their previous classification as ellipticals.  We
follow Graham (2002a) in classifying M32 as a probable stripped S0, in
which the bulge resides within a remnant exponential disk/envelope.

 %
 %

\section{Results and Discussion}

\begin{figure}
  \centering
\includegraphics[scale=0.6]{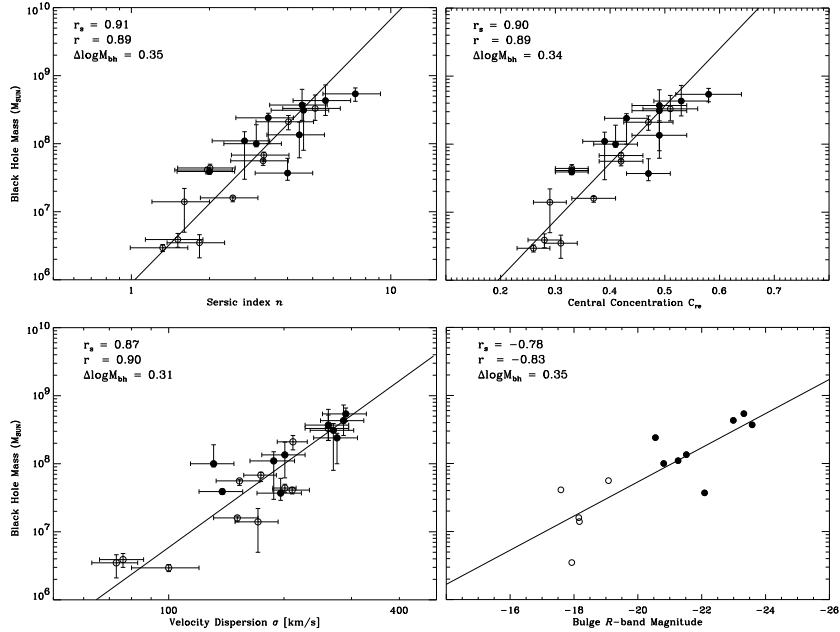}
  \caption{Correlations between SMBH mass $M_{\rm BH}$ and: S\'ersic
  index $n$ (upper left), central concentration $C_{re}$ (upper
  right), central velocity dispersion (lower left), and bulge $R$-band
  luminosity (lower right).  The straight lines are fits made with the
  bisector linear-regression routine from Akritas \& Bershady (1996). 
  We also show the Spearman rank-order correlation coefficient
  $r_{s}$ and the Pearson linear correlation coefficient $r$.  (The
  Spearman coefficient is more robust to outliers and does not
  presuppose a linear relation.)  Filled circles are elliptical
  galaxies; open circles are bulges of disk galaxies.}
  \label{fig:erwin-correlations}
\end{figure}

We find a strong correlation between the central SMBH mass $M_{\rm
BH}$ and its host galaxy's bulge structure, as measured by the
S\'ersic index $n$ (or by central concentration $C_{re}$; Graham et
al.\ 2001b), such that more centrally concentrated galaxies (higher
$n$) have more massive black holes.  This correlation is as good as
that previously found between $M_{\rm BH}$ and the central velocity
dispersion, when the same galaxies are compared
(Figure~\ref{fig:erwin-correlations}).  In addition, by identifying S0
galaxies misclassified as ellipticals and making careful bulge-disk
decompositions --- including the avoidance of bars and the
accomodation of other components such as inner disks --- we find bulge
luminosities for these galaxies which also correlate well with SMBH
mass (McLure \& Dunlop 2002 found a similary tight correlation by
considering only elliptcal galaxies.).  The scatter in $\log(M_{\rm
BH})$ for these three relations is comparable (0.31Ð-0.35 dex).

We thus have four closely-linked quantities --- $M_{\rm BH}$, central
velocity dispersion, bulge luminosity, and global bulge structure ---
which are well correlated at least over the range $M_{\rm BH} \sim
10^{6}$-Ð$10^{9} M_{\odot}$.  Models which explain SMBH formation
and growth ultimately need to address all three SMBH-bulge relations;
explanations which rely primarly on one quantity (e.g., bulge mass or
central velocity dispersion) need to explain why the other quantities
correlate so well.

It is worth noting that using the $n$-Ð$M_{\rm BH}$ relation to
estimate SMBH masses --- or to study SMBH-bulge evolution with
redshift --- has several advantages over the $\sigma_{0}$-Ð$M_{\rm
BH}$ and $L_{\rm bulge}$-Ð$M_{\rm BH}$ relations.  Measurements of
$n$ require only (uncalibrated) images, and are thus less expensive in
terms of telescope time than the spectroscopic observations needed to
determine $\sigma_{0}$.  Measurements of $n$ are also not sensitive to
uncertainties in, e.g., distance or Galactic extinction.

Finally, the existence of a clear relation between $n$ and SMBH mass
is further proof that bulges are \textit{not} homologous: not all
ellipticals have de Vaucouleurs ($n = 4$) profiles, and spiral and S0
bulges cannot simply be classified as either de Vaucouleurs or
exponential ($n = 1$).



\end{document}